\documentclass[notitlepage, twocolumn]{revtex4-1}

\usepackage[T1]{fontenc}
\usepackage{graphicx}
\usepackage{bm}
\usepackage{amsmath}
\usepackage{times}
\usepackage[version=3]{mhchem}

\newcommand{\vect}[1]{\bm{#1}}
\newcommand{\tens}[1]{\mathbf{#1}}

\newcommand{\icm}{cm$^{-1}$}

\usepackage{xcolor}

\begin{document}

\title{Unifying Low-Frequency Raman and Infrared Signatures across Perovskites}

\author{Ethan Berger}
\email{ethan.berger@chalmers.se}
\affiliation{Department of Physics and Astronomy, Chalmers University of Technology, SE-41296 Gothenburg, Sweden}
\author{Erik Fransson}
\affiliation{Department of Physics and Astronomy, Chalmers University of Technology, SE-41296 Gothenburg, Sweden}
\author{Apinya Ngoipala}
\affiliation{Department of Physics and Astronomy, Chalmers University of Technology, SE-41296 Gothenburg, Sweden}
\author{Lucas Svensson}
\affiliation{Department of Physics and Astronomy, Chalmers University of Technology, SE-41296 Gothenburg, Sweden}
\author{Priyanka Deswal}
\affiliation{Department of Physics and Astronomy, Chalmers University of Technology, SE-41296 Gothenburg, Sweden}
\author{Mehmet Baskurt}
\affiliation{Department of Physics and Astronomy, Chalmers University of Technology, SE-41296 Gothenburg, Sweden}
\author{Paul Erhart}
\affiliation{Department of Physics and Astronomy, Chalmers University of Technology, SE-41296 Gothenburg, Sweden}
\author{Julia Wiktor}
\email{julia.wiktor@chalmers.se}
\affiliation{Department of Physics and Astronomy, Chalmers University of Technology, SE-41296 Gothenburg, Sweden}

\begin{abstract}
Perovskites have been studied for a wide range of applications, from oxide-based ferroelectrics to halide-perovskite photovoltaics. 
Despite their diverse functionalities, these materials share strong lattice anharmonicity arising from structural instabilities such as polar off-centering and octahedral tilting.
These anharmonic distortions strongly influence the low-frequency vibrational response and can be observed in infrared (IR) and Raman spectra.
However, low-frequency IR and Raman signatures are typically investigated separately, hindering the development of a unified understanding of vibrational spectroscopy across perovskite classes.
In this work, we use machine-learning-based molecular dynamics to simulate the IR and Raman spectra of oxide perovskites (\ce{SrTiO3} and \ce{BaTiO3}), halide perovskites (\ce{CsPbBr3}, \ce{CsSnBr3}, and \ce{CsGeBr3}) and double halide perovskites (\ce{Cs2AgAlBr6} and \ce{Cs2AgBiBr6}).
By comparing materials exhibiting polar, tilt-driven and mixed instabilities, we find that polar off-centering leads to enhanced low-frequency IR and dielectric responses, whereas nonpolar modes, particularly octahedral tilting, are associated with Raman central peaks.
The strength of the low-frequency features also depends on how strongly the underlying modes couple to polarization or dielectric susceptibility.
Finally, we show that the dynamics of these instabilities play an important role, with slow polar reorientations shifting the dielectric response to very low frequencies and overdamped tilting modes producing persistent Raman central peaks.
These results establish a unified framework connecting low-frequency IR and Raman spectra to the symmetry, coupling and dynamics of their underlying anharmonic lattice instabilities.
\end{abstract}

\maketitle

%\begin{figure}[t]
%    \centering
%    \includegraphics[width=3.3in]{TOC.pdf}
%\end{figure}

% ============================================================
\section{Introduction}

Perovskite materials with stoichiometry \ce{ABX3} constitute one of the largest and most versatile families of functional materials. 
Their exceptional chemical flexibility has enabled the development of various fields, from oxide ferroelectrics and piezoelectrics \cite{Rodel2009_perspective,Humayun2025_ABO3_review,Basith2025_piezo_review,Grinberg2013_oxides,Bai2015_BCZT,Liu2024_PiezoBiFeO3} to the more recent applications of halide perovskites in solar cells and optoelectronic devices \cite{Green2014_perovskite_review,Sutherland2016_perovskite_review,Kovalenko2017_optoelectronic,Liu2021_perovskite_review,Berger2022_scalable,Yang2026_review,Wang2023_nature}. 
Beyond conventional oxide and halide compounds, the perovskite family also encompasses double perovskites \cite{Igbari2019_double,Chu2019_double,Lei2021_double,Filip2016_double}, chalcogenide perovskites \cite{Sun2015_chalcogenide,Tiwari2021_chalcogenide, Sopiha2022_chalcogenide,Nishigaki2020_chalcogenide,Kayastha2025_chalcogenide}, and low-dimensional perovskites \cite{Yusoff2018_lowD,Azmi2026_lowD,Mao2018_lowD,Kayastha2026_lowD_chalcogenide}, further expanding the range of accessible functionalities.
Alloying and doping offer additional avenues for tuning the properties of existing perovskite materials \cite{Zhou2016_doping_review,Ono2017_alloying_review,Li2020_alloying,Zhang2023_alloying,Fu2018_doping,Zhao2019_doping}.
This structural and chemical versatility provides an exceptionally broad playground for tailoring physical properties through composition and crystal chemistry.
In addition to this remarkable diversity, perovskites generally possess highly anharmonic lattices that give rise to unusual behaviors at finite temperatures and rich phase diagrams \cite{Munson2019_anharmonicity_perspective,Pertsev1998_diagrams,Zhong1995_STO,Hainer2025_morphotropic}. 
In most perovskites, this anharmonicity originates from a small number of lattice instabilities, primarily tilting of the \ce{BX6} octahedra and polar off-centering of the B site \cite{Yang2017_spontaneous, Yang2020_instabilities,Gigli2022_BTO,Zhong1995_STO}. 
These distortions govern structural phase transitions and strongly influence dielectric, ferroelectric, optical, and thermal properties. 
Establishing general relationships between these structural instabilities and experimentally accessible observables is therefore essential for understanding and designing functional perovskite materials.

\begin{figure*}
    \centering
    \includegraphics[width=7in]{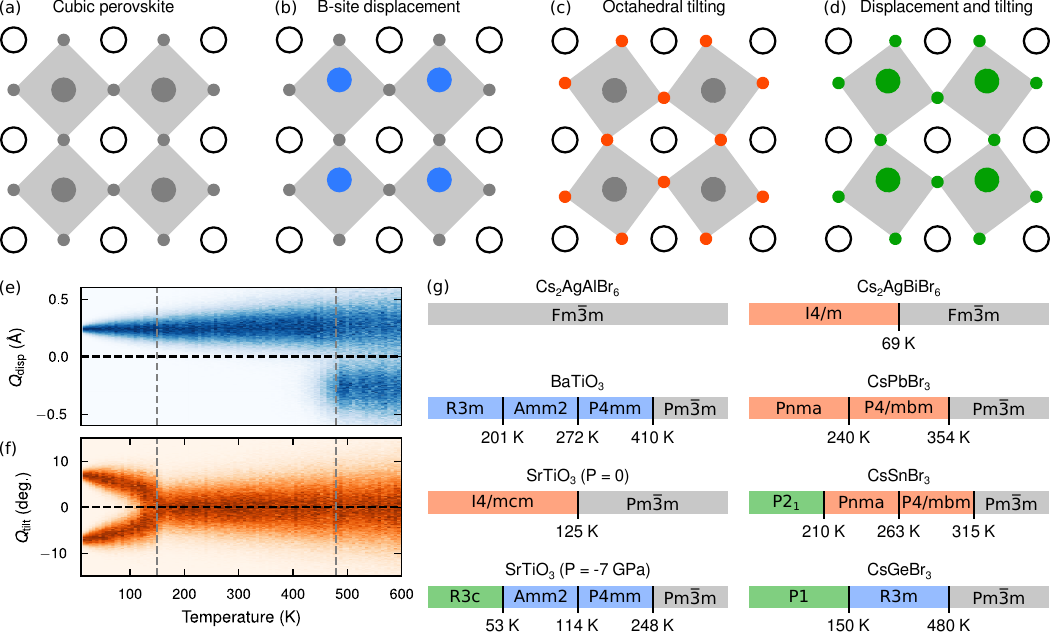}
    \caption{Structural instabilities and phase behavior across the investigated perovskites. 
    (a) High-symmetry cubic perovskite structure.
    (b–d) Representative distortions associated with B-site off-centering (b), octahedral tilting (c), and their combination (d).
    (e,f) Temperature-dependent distributions of the B-site off-centering $Q_\text{disp}$ (e) and octahedral tilting $Q_\text{tilt}$ (f) for \ce{CsGeBr3}. 
    Vertical gray lines highlight the phase transition temperatures. 
    (g) Summary of the phases and transition temperatures of the investigated perovskites. 
    Blue, orange, and green denote phases characterized by off-centering, octahedral tilting, and both distortions, respectively, while gray denotes the cubic phase.
    }
    \label{fig:Fig1}
\end{figure*}

Among these lattice instabilities, polar off-centering in oxide perovskites is well known to give rise to soft phonon modes.
Near the phase transition, the frequency of these modes vanishes, leading to a substantial enhancement of the dielectric response.
This enhancement is central to the design of high-performance piezoelectric materials, where alloying is commonly employed to tune structural instabilities and maximize the dielectric response.
A more modest dielectric enhancement has also been observed in hybrid halide perovskites and attributed to a combined motion between the organic molecules and the inorganic octahedra \cite{Berger2020_MAPI, Herz2018_JPCL, Anusca2017_AdvEM}.
Another prominent low-frequency feature is the central peak observed in Raman spectra of oxide perovskites near structural phase transitions \cite{Lyons1977_central,Fontana1988_central}.
Although two-phonon processes can produce a central Raman contribution in crystals \cite{Lyons1980_xenon}, the pronounced intensity and strong temperature dependence of central peaks in perovskites have raised debate over their origin \cite{Menahem2023_PRM,Rosander2025_Raman}.
In the ideal cubic phase (Pm$\bar{3}$m), such peaks cannot be explained by first-order Raman scattering, which is forbidden by symmetry.
Raman central peaks have also been observed in hybrid and inorganic halide perovskites, indicating that they cannot originate solely from the rotation of the organic molecules \cite{Yaffe2017_PRL}.
Machine-learning-based molecular dynamics simulations further linked the central peak to dynamic octahedral tilting \cite{Berger2024_Polarizability}.
However, recent work compared the Raman spectra of \ce{CsPbBr3} and \ce{Cs2AgBiBr6}, and showed that the latter exhibit no persistent central peak at high temperatures, despite the presence of anharmonic tilting modes \cite{Cohen2022_AdvMater}.
More recently, a study compared the low-frequency Raman spectra of various halide perovskites, arguing that the central peak cannot originate from octahedral tilting, but instead from broad Raman-active modes amplified by Bose-Einstein statistics at low frequencies \cite{Lim2024_ACSEnerLett}.
While the relationship between polar soft modes and dielectric enhancement is well established, the microscopic origin of Raman central peaks remains debated.
Moreover, the low-frequency IR and Raman signatures are generally interpreted independently and often within separate material communities.
Consequently, a unified understanding of vibrational spectra and their connection to the underlying lattice instabilities across different perovskite classes remains lacking.

In this work, we use machine-learning-based molecular dynamics to investigate the Raman and infrared spectra of representative oxides (\ce{BaTiO3} and \ce{SrTiO3}), halides (\ce{CsPbBr3}, \ce{CsSnBr3} and \ce{CsGeBr3}), and double perovskites (\ce{Cs2AgAlBr6} and \ce{Cs2AgBiBr6}).
By systematically comparing materials exhibiting octahedral tilting, polar off-centering, and mixed structural instabilities, we establish a unified framework relating low-frequency vibrational signatures to the underlying anharmonic modes.
To quantify these signatures, we use the static dielectric constant and the intensity of the Raman central peak as descriptors of the infrared and Raman responses, respectively.
%We show that both quantities reach a maximum near structural phase transitions, providing a direct connection between the low-temperature distorted phases and the vibrational response of the high-temperature cubic phase. 
By following the vibrational responses across the phase transitions, we show that the low-frequency spectra are connected to the anharmonic instabilities responsible for the low-temperature distorted phases.
Finally, we discuss the conditions under which these spectroscopic signatures emerge, highlighting the roles of infrared/Raman mode couplings and damping in determining the vibrational response of perovskite materials.

% ============================================================
\section{Results and Discussion}

% ------------------------------------------------------------
\subsection{Structural landscape and phase transitions}

\begin{figure*}
    \centering
    \includegraphics[width=7in]{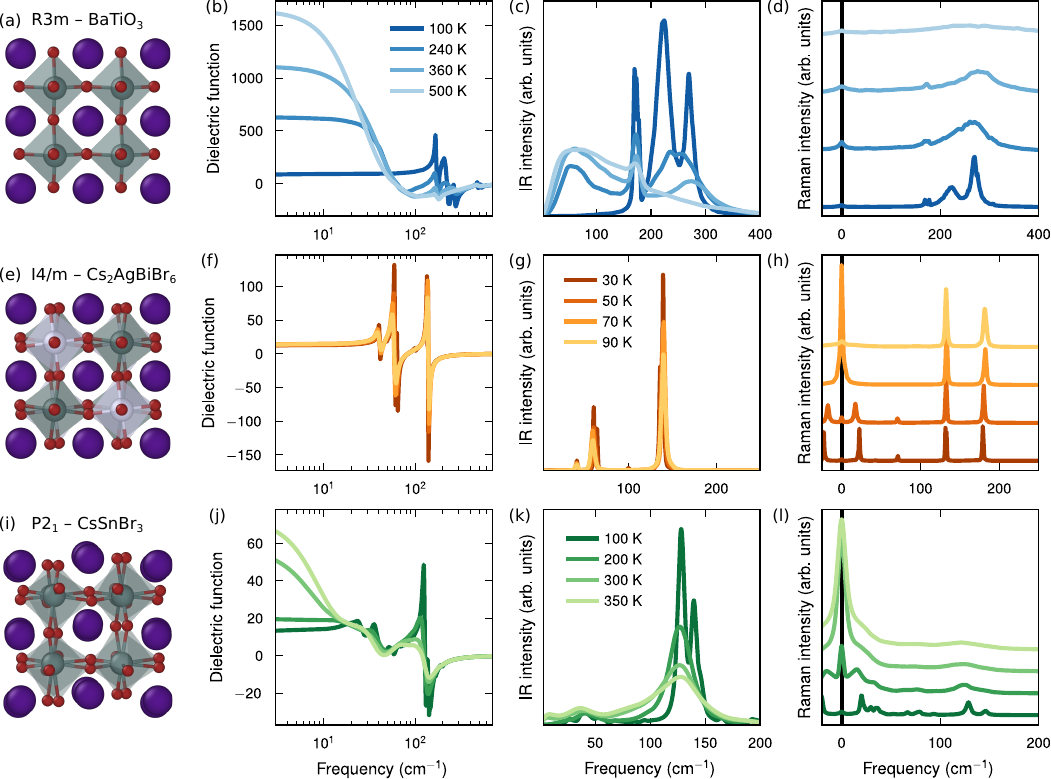}
    \caption{Representative dielectric, infrared, and Raman responses for different structural instabilities. 
    (a,e,i) Ground-state structures of \ce{BaTiO3} (a), \ce{Cs2AgBiBr6} (e), and \ce{CsSnBr3} (i). 
    A-site cations are in purple, B-site metals in gray (and white in \ce{Cs2AgBiBr6}), and oxygen/bromine in dark red.
    (b,f,j) Real part of the dielectric function, (c,g,k) infrared spectra, and (d,h,l) Raman spectra of the corresponding materials at selected temperatures.
    Temperatures are chosen to sample the distinct structural phases of each material.
    }
    \label{fig:Fig2}
\end{figure*}

The structural instabilities considered in this work are illustrated in Figure~\ref{fig:Fig1}(a-d).
Starting from the high-symmetry cubic structure (a), the dominant distortions are B-site off-centering (b), octahedral tilting (c), or a combination of both (d).
To characterize their evolution with temperature, we perform heating simulations using charge-informed neuroevolution potentials (qNEPs) \cite{Fan2026_qnep} and monitor the lattice parameters together with order parameters for B-site off-centering $Q_\text{disp}$ and octahedral tilting $Q_\text{tilt}$.
Further details regarding the training of the qNEP models and the heating simulations are provided in the Methods and Supporting Information.
Figures~\ref{fig:Fig1}(e,f) illustrate an example for \ce{CsGeBr3}, where both types of instability are present in the low-temperature phase. 
The first phase transition can be identified at 150~K from the evolution of $Q_\text{tilt}$, while $Q_\text{disp}$ shows the second phase transition at 480~K. 
As a result, \ce{CsGeBr3} evolves from a P1 ground state containing both off-centering and octahedral tilting to polar R3m phase, and finally cubic Pm$\bar{3}$m phases. 
Corresponding order parameters for the remaining materials are provided in the Supporting Information.

A similar analysis is repeated for 7 different perovskites, and the resulting structural landscape is summarized in Figure~\ref{fig:Fig1}(g).
The materials are specifically selected to span distinct classes of lattice instability. 
First, \ce{Cs2AgAlBr6} remains cubic and provides a reference without any structural instability.
For tilting, \ce{CsPbBr3} is selected as a reference material, as its anharmonic behavior has been extensively studied \cite{Stoumpos2013_CsPbBr3,Wiktor2023_CsPbBr3,Fransson2023_CsPbBr3}. 
We also include \ce{Cs2AgBiBr6}, which exhibits similar octahedral-tilting instabilities \cite{Baskurt2026_Faraday,Tower_Tc_AgBi} but shows no Raman central peak at high temperatures \cite{Cohen2022_AdvMater}. 
In contrast, \ce{BaTiO3} provides the canonical polar reference, undergoing successive off-centering-driven transitions before reaching its cubic phase \cite{Merz1949_BTO,Wieder1955_BTO,Hashimoto2015_BTO,Gigli2022_BTO}. 
\ce{CsSnBr3} exhibits both off-centering and octahedral tilting, thereby providing a mixed case in which the two spectroscopic channels can coexist \cite{Mori1986_Tc_CsSnBr3,Fabini2024_Tc_CsSnBr3}.
Experimental reports show that \ce{CsGeBr3} has an R3m to cubic transition at high temperature \cite{Thiele1987_CsGeBr3}, making it different from \ce{CsPbBr3} and \ce{CsSnBr3}.
Additionally, recent DFT calculations and machine-learning-based molecular dynamics revealed a ground state with both kinds of instabilities, leading to a second phase transition at low temperature \cite{Baskurt2026_Ge,Deswal2026_Ge}.
Finally, \ce{SrTiO3} provides a tunable comparison: at zero pressure its ground state contains only octahedral tilting, whereas applying negative pressure induces additional off-centering \cite{Zhong1995_STO,Guennou2010_STO,Nova2019_STO}. 
A summary of the calculated structural properties is provided in Table~S3 of the Supporting Information, showing good agreement with both previous theoretical models and available experimental results.
Together, these systems provide a controlled structural basis for examining how polar and nonpolar instabilities manifest in the low-frequency infrared and Raman response.

% ------------------------------------------------------------
\subsection{Spectroscopic fingerprints of low-frequency dynamics}

We start by examining the infrared and Raman response of three representative materials, \ce{BaTiO3}, \ce{Cs2AgBiBr6}, and \ce{CsSnBr3}, which provide examples of off-centering, octahedral-tilting, and mixed structural instabilities, respectively. 
Complete spectra for all investigated materials are provided in Figures~S1--S8 of the SI. 
The procedures used to calculate the dielectric function and the infrared and Raman spectra, together with the molecular dynamics parameters, are described in detail in the Methods section.

In its ground state, \ce{BaTiO3} exhibits B-site off-centering along all three pseudocubic directions, resulting in the rhombohedral R3m structure shown in Figure~\ref{fig:Fig2}(a). 
The successive phase transitions can be understood as the progressive loss of components of this off-centering, leading to the sequence R3m $\rightarrow$ Amm2 $\rightarrow$ P4mm $\rightarrow$ Pm$\bar{3}$m, similar to that obtained for \ce{SrTiO3} under negative pressure in Figure~\ref{fig:Fig1}(g). 
The low-frequency dielectric response is strongly enhanced across the successive phases, as seen in Figure~\ref{fig:Fig2}(b). 
This enhancement is associated with the softening of polar modes, which is directly visible in the infrared spectra in Figure~\ref{fig:Fig2}(c). 
At 100~K, the IR spectrum is dominated by relatively sharp modes between 180 and 300~\icm. 
With increasing temperature, these modes soften and broaden, while spectral weight progressively shifts toward frequencies below 100~\icm, such that the low-frequency response dominates in the cubic phase at 500~K. 
This behavior is characteristic of displacive ferroelectrics and is also observed for \ce{SrTiO3} (SI Figures~S7 and S8).

A different behavior is observed for the double perovskite \ce{Cs2AgBiBr6}, whose low-temperature structure is characterized by octahedral tilting without B-site off-centering and adopts the I4/m space group represented in Figure~\ref{fig:Fig2}(e). 
In contrast to \ce{BaTiO3}, the dielectric function presented in Figure~\ref{fig:Fig2}(f) shows no pronounced low-frequency enhancement across the temperatures considered.
This is consistent with the infrared spectra retaining finite-frequency features whose positions remain approximately unchanged with temperature, as can be seen in Figure~\ref{fig:Fig2}(g). 
The strongest temperature dependence instead occurs in the Raman spectrum visible in Figure~\ref{fig:Fig2}(h).
At low temperature, the Raman spectrum shows three clear peaks at 22, 132 and 179~\icm, as well as a fainter peak at 71~\icm.
Upon heating, the low-frequency mode associated with octahedral tilting progressively softens and develops into an intense central peak close to the transition to the cubic phase around 70~K.
Above the transition, the central peak disappears, while the two higher-frequency Raman modes remain, as allowed by the ordered double-perovskite structure (space group Fm$\bar{3}$m).
Similar modes are also found in cubic \ce{Cs2AgAlBr6}, which, in contrast, shows no central peak because it lacks a soft octahedral-tilting instability (Figure~S1). 
The series of peaks at low and high temperatures, together with the softening of the tilting mode in \ce{Cs2AgBiBr6} have also been observed experimentally \cite{Cohen2022_AdvMater}, although no central peak was reported.
This double perovskite thus provides an example of an anharmonic structural instability that produces a pronounced central Raman response without a corresponding dielectric enhancement, in strong contrast with \ce{BaTiO3}.

Finally, \ce{CsSnBr3} combines B-site off-centering and octahedral tilting in its low-temperature structure, leading to the P2$_1$ ground state shown in Figure~\ref{fig:Fig2}(i) \cite{Mori1986_Tc_CsSnBr3,Fabini2024_Tc_CsSnBr3}.
Correspondingly, signatures of both types of instability are present in its spectroscopic response. 
In Figure~\ref{fig:Fig2}(j), the dielectric function exhibits an enhanced low-frequency response associated with the polar character of the material, although its magnitude remains considerably smaller than in the oxide ferroelectric \ce{BaTiO3}. 
In contrast to \ce{BaTiO3}, however, the infrared spectra shown in Figure~\ref{fig:Fig2}(k) do not exhibit a clear soft-mode evolution with temperature. 
The Raman response is more similar to that of \ce{Cs2AgBiBr6}, as can be seen in Figure~\ref{fig:Fig2}(l). 
At low temperature, the spectrum shows a few sharp peaks, which progressively soften upon heating and merge into a pronounced central peak around the structural phase transition. 
In contrast to \ce{Cs2AgBiBr6}, this central peak remains strong above the transition and eventually dominates the high-temperature Raman spectrum. 
A similar behavior is also found for \ce{CsPbBr3} (see SI Figure~S3), and has already been highlighted in previous simulations \cite{Berger2024_Polarizability}, as well as observed experimentally for \ce{CsPbBr3} and methylammonium lead bromide \cite{Yaffe2017_PRL}.
\ce{CsSnBr3} therefore provides a mixed case in which pronounced low-frequency responses are observed in both infrared and Raman spectroscopy.

% ------------------------------------------------------------
\subsection{A unified Raman--IR classification}

\begin{figure*}
    \centering
    \includegraphics[width=7in]{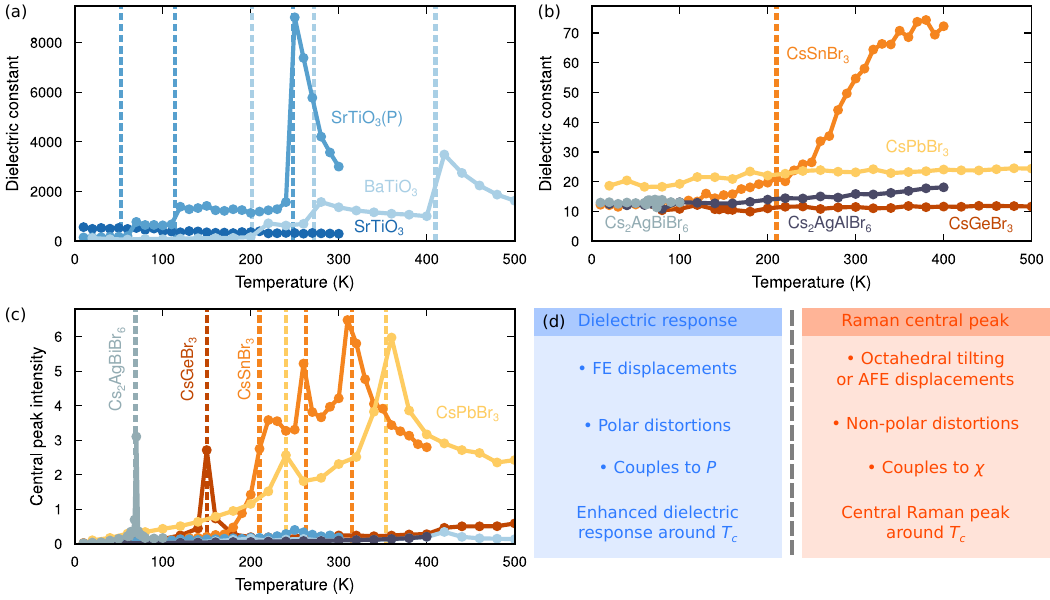}
    \caption{
    Temperature dependence of low-frequency infrared and Raman descriptors. 
    (a,b) Static dielectric constant for oxide (a) and halide (b) perovskites. 
    (c) Raman central-peak intensity for the investigated materials. 
    Vertical dashed lines indicate structural phase-transition temperatures. 
    (d) Summary of the structural, symmetry, and spectroscopic characteristics associated with off-centering and octahedral-tilting instabilities.
    }
    \label{fig:Fig3}
\end{figure*} 

To compare the low-frequency infrared and Raman spectra across the complete set of materials, we need to define a representative descriptor for each response.
For the infrared, the static dielectric constant provides a natural descriptor and is shown for the oxide perovskites in Figure~\ref{fig:Fig3}(a).
Both \ce{BaTiO3} and \ce{SrTiO3} under negative pressure show a large enhancement of the dielectric response around their phase transitions, which are highlighted with vertical dashed lines. 
Both materials exhibit pronounced off-centering in their low-temperature phases, whereas \ce{SrTiO3} at zero pressure does not.
Correspondingly, its dielectric response is significantly lower than that of its two counterparts that highlighted B-site off-centering.
A similar observation can be made for halide perovskites, presented in Figure~\ref{fig:Fig3}(b).
Most halides considered here show no anharmonic B-site off-centering, and therefore show no significant enhancement of their dielectric response upon phase transition.
\ce{CsSnBr3}, which exhibits B-site off-centering, follows this trend and shows an enhanced dielectric response. 
By contrast, \ce{CsGeBr3} shows no comparable enhancement despite retaining ordered off-centering up to 480~K.
Overall, the comparison reveals a strong association between polar off-centering and enhanced dielectric response, with \ce{CsGeBr3} providing a notable exception.

While the connection between polar instabilities and dielectric enhancement is well established, the microscopic origin of the low-frequency Raman response remains more debated.
For the low-frequency Raman descriptor, we simply choose to use the intensity of the central peak, normalized by the maximal intensity of finite-frequency peaks at low temperatures.
Figure~\ref{fig:Fig3}(c) shows the central peak intensity as a function of temperature for all materials.
Most halide perovskites show large central peak intensities, with maxima located at their phase transition temperatures (represented by vertical dashed lines). 
From Figure~\ref{fig:Fig1}(g), most phase transitions leading to an increase of the Raman central peak are related to the octahedral tilting.
The only notable exception is the first phase transition of \ce{CsSnBr3} at 210~K, which is related to B-site off-centering and is also found to lead to a faint increase in the Raman central peak intensity.
In contrast, \ce{BaTiO3}, which only exhibits off-centering modes, does not show any central Raman peak.
This difference could be explained by the off-centering arrangement being ferroelectric (FE) in \ce{BaTiO3} but partly anti-ferroelectric (AFE) in \ce{CsSnBr3} (see Figure~\ref{fig:Fig5}).
Finally, we note that \ce{SrTiO3} also does not show a notable central Raman feature, although it shows octahedral tilting at both pressures considered. 

By combining results from Figure~\ref{fig:Fig3}(a-c), it becomes clear that the formation of a Raman central peak in perovskites is closely related to phase transitions and the symmetry of the underlying instabilities. 
At low temperatures, these instabilities give rise to finite-frequency vibrational features in the low-symmetry phases. 
Upon approaching a phase transition, their frequency approaches zero, leading to an increase of the associated low-frequency spectra.
Ferroelectric displacement of B-sites represents a polar mode and therefore couples directly to the infrared response and the dielectric function.
On the other hand, tilting of the \ce{BX6} octahedra and AFE off-centering are nonpolar distortions and are found here to contribute to the low-frequency Raman response.
%This correspondence echoes the usual selection rules of vibrational spectroscopy. 
The resulting symmetry-based classification is summarized in Figure~\ref{fig:Fig3}(d).

Although most perovskites follow the proposed mechanism, there remain open questions regarding Figures~\ref{fig:Fig3}(a-c). 
As previously mentioned, \ce{CsGeBr3} and \ce{SrTiO3} should show low-frequency responses that we are missing here.
%Similarly, the first phase transition of \ce{CsSnBr3} relates to off-centering and still shows a Raman response, in contradiction with the selection rule.
Looking at the Raman responses in Figure~\ref{fig:Fig3}(c), \ce{Cs2AgBiBr6} and \ce{CsGeBr3} do not retain their central peak intensity after phase transition, while the Raman spectra of \ce{CsPbBr3} and \ce{CsSnBr3} are dominated by the central peak in the cubic phase.
%Through the investigation of the static coupling and the dynamics of the anharmonic mode, we attempt at explaining some of the unique features observed in Figure~\ref{fig:Fig3}.
To clarify the origin of these material-dependent features, we therefore examine the static coupling and finite-temperature dynamics of the relevant anharmonic modes.

% ------------------------------------------------------------
\subsection{Static coupling of anharmonic modes}

\begin{figure*}
    \centering
    \includegraphics[width=7in]{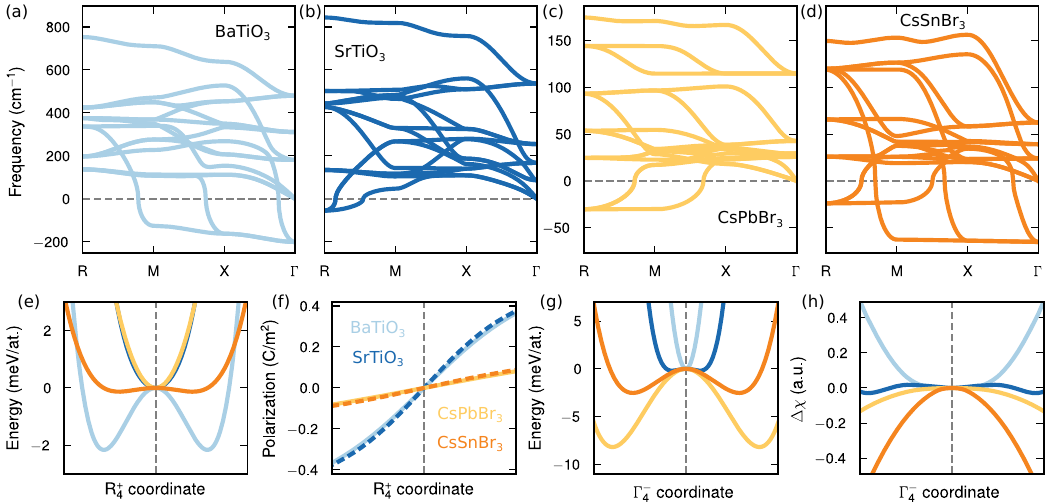}
    \caption{
    Static coupling between structural instabilities and spectroscopic observables. 
    (a–d) Phonon dispersions of cubic \ce{BaTiO3} (a), \ce{SrTiO3} at zero pressure (b), \ce{CsPbBr3} (c), and \ce{CsSnBr3} (d), showing polar and/or octahedral-tilting instabilities. 
    (e,f) Potential energy (e) and polarization (f) along the $\Gamma_4^-$ off-centering instability. 
    (g,h) Potential energy (g) and variation of the dielectric susceptibility (h) along the R$_4^+$ octahedral-tilting instability.
    Vertical dashed lines indicate the high-symmetry cubic structure.
    }
    \label{fig:Fig4}
\end{figure*}

Next, we investigate the static dielectric and Raman coupling via phonon modes.
Such static mode projections can provide a simplified picture of the symmetry and strength of the coupling between each structural distortion and the corresponding spectroscopic observable. 
Figure~\ref{fig:Fig4}(a-d) shows the phonon dispersion curves of cubic \ce{BaTiO3}, \ce{SrTiO3}, \ce{CsPbBr3}, and \ce{CsSnBr3}, respectively.
The instabilities identified through heating and presented in Figure~\ref{fig:Fig1}(g) are represented by imaginary modes in the phonon dispersion curves.
For example, B-site off-centering modes are visible at $M$, $X$, and $\Gamma$ for \ce{BaTiO3}.
Imaginary modes of \ce{CsPbBr3} at $R$ and $M$ instead correspond to out-of-phase and in-phase octahedral tilt modes.
A similar out-of-phase $R$ mode is also visible in the phonon dispersion curve of \ce{SrTiO3}.
Finally, \ce{CsSnBr3} shows a combination of the off-centering modes at $M$, $X$, and $\Gamma$ seen in \ce{BaTiO3} and tilting modes at $R$ and $M$ seen in \ce{CsPbBr3}, in agreement with the fact that this material shows both kinds of instabilities in its ground state.
These results agree well with the phase diagrams obtained using MD heating runs and presented in Figure~\ref{fig:Fig1}(g).

We focus here on the $\Gamma_4^-$ off-centering mode and the $R_4^+$ tilting mode, and investigate their coupling to the infrared and Raman spectra. 
Other off-centering ($X_5^-$, and $M_2^-$) and tilting ($M_3^+$) modes are presented in more details in SI Figure~S10.
Figure~\ref{fig:Fig4}(e) shows the energy along the $\Gamma_4^-$ mode, exhibiting a typical double-well potential for \ce{BaTiO3} and \ce{CsSnBr3}.
The polarization along the same path is then presented in Figure~\ref{fig:Fig4}(f), with oxide perovskites showing larger polarization than halide ones.
This difference can be explained by the larger oxidation state and Born effective charges present in the oxides, as well as by their smaller volume.
The infrared intensity of a mode is proportional to the variation of the polarization $\left|\frac{dP}{dQ}\right|^2$.
Here, it is clear that oxides have a higher infrared intensity, which agrees well with our results in Figure~\ref{fig:Fig3}(a-b). 

Figure~\ref{fig:Fig4}(g) then shows the potential energy along the tilt mode $R_4^+$, while Figure~\ref{fig:Fig4}(h) shows the variation of the dielectric susceptibility $\Delta \chi$ along the same path.
Here again, typical double-well potentials are recovered for materials with imaginary modes, that is \ce{SrTiO3}, \ce{CsPbBr3}, and \ce{CsSnBr3}.
The linear Raman coupling is determined by the variation of the dielectric susceptibility along a given mode, $\frac{d\chi}{dQ}$.
For these cubic tilt modes, the susceptibility is symmetric around $Q=0$, such that $\frac{d\chi}{dQ}=0$ at $Q=0$, and the linear Raman coupling vanishes.
%This is typical of non-$\Gamma$ modes, for which the first order Raman tensor is always zero.
In contrast, for the polar off-centering mode considered above, the infrared coupling $\frac{dP}{dQ}$ remains finite at $Q=0$. 
This difference provides a simple static picture for why the dielectric response remains enhanced above the phase transition, while the Raman central response is strongest around the transition and tends to decrease in the cubic phase. 
As discussed below, the finite-temperature dynamics can modify this picture and allow a central Raman feature to persist in the cubic phase.

One can also note that for \ce{SrTiO3}, the local minima of the double-well are much closer to $Q=0$ than for halides. 
Due to its quadratic shape, the variation $\frac{d\chi}{dQ}$ remains small close to $Q=0$, especially for \ce{SrTiO3}.
Consequently, the Raman intensity of \ce{SrTiO3} is much weaker than in other materials, as can be seen from Figure~S9 of the SI.
This low intensity can explain the absence of a clear Raman central peak in \ce{SrTiO3} previously observed in Figure~\ref{fig:Fig3}(c). 

\begin{figure*}
    \centering
    \includegraphics[width=7in]{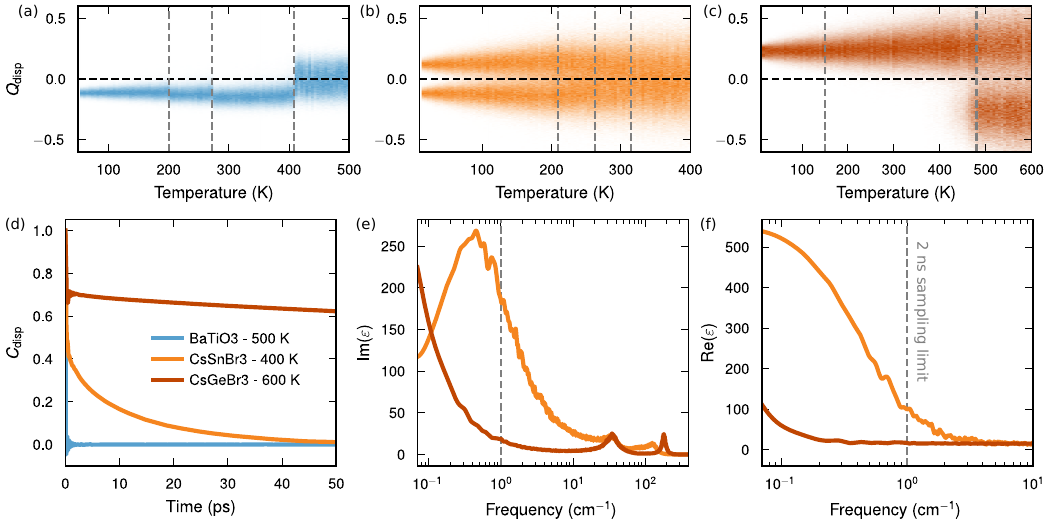}
    \caption{
    Dynamics and ordering of the B-site off-centering.
    (a-c) Off-centering along the Y axis during heating for (a) \ce{BaTiO3}, (b) \ce{CsSnBr3}, and (c) \ce{CsGeBr3}.
    Phase transition temperatures are highlighted by vertical dashed lines.
    (d) Off-centering autocorrelation function $C_{\mathrm{disp}}(t)$ for the same materials in the cubic phase.
    (e-f) Imaginary and real parts of the dielectric function of \ce{CsSnBr3} and \ce{CsGeBr3} from 50 ns simulations.
    The lowest sampled frequency from 2 ns simulations is highlighted by a vertical dashed line.
    }
    \label{fig:Fig5}
\end{figure*}

Overall, these results show that the appearance of low-frequency vibrational responses depends not only on the symmetry of the instability, but also on its coupling to the spectroscopic observable. 
However, this static picture remains incomplete, as it does not account for the finite-temperature dynamics of the anharmonic modes. 
The resulting spectral response can also depend on the characteristic dynamics of these modes, including their reorientation and damping.

% ------------------------------------------------------------
\subsection{Dynamic origin of low-frequency responses}

So far, we only considered static properties, such as the symmetry and the coupling, but the vibrational response at finite temperatures also depends on the dynamics of the active modes.

We first investigate the dynamics of off-centering modes in \ce{BaTiO3}, \ce{CsSnBr3}, and \ce{CsGeBr3}.
The off-centering of their B-site atoms during heating is shown in Figures~\ref{fig:Fig5}(a-c), with phase transitions highlighted by vertical dashed lines.
The off-centering in \ce{BaTiO3} evolves from a ferroelectric alignment to a paraelectric regime, with each phase transition corresponding to one direction changing regime.
In contrast, \ce{CsSnBr3} and \ce{CsGeBr3} are found to move from polar ground states to an order-disorder phase with very long reorientation times \cite{Fransson2026_Ge}.
The latter are represented in Figure~\ref{fig:Fig5}(d), showing time scales of less than 1 ps for \ce{BaTiO3}, but of the order of 50 ps for \ce{CsSnBr3} and much longer for \ce{CsGeBr3}.
These long reorientation timescales correspond to frequencies below 1~\icm, below the range typically sampled by infrared or Raman spectroscopy \cite{Koch2023_LFIR,Remoto2026_LFR}.
Additionally, the simulation time of 2 ns used in this work may not provide sufficient sampling of these slow dynamics, particularly for \ce{CsGeBr3}.
To better resolve these low-frequency responses, Figures~\ref{fig:Fig5}(e-f) show the imaginary and real parts of the dielectric function for \ce{CsSnBr3} and \ce{CsGeBr3} in their cubic phase using a total simulation length of 50 ns.
The lowest frequency sampled in the 2 ns simulations is highlighted by a vertical line, clearly showing that a longer simulation length is necessary to properly sample these slow fluctuations.
For \ce{CsSnBr3}, a simulation length of 2 ns already showed part of the dielectric enhancement as observed in Figure~\ref{fig:Fig3}(b), but did not fully capture the response. 
With 50 ns, the peak of the imaginary part is almost completely resolved, and the real part starts to reach a plateau.
On the contrary, the dielectric response of \ce{CsGeBr3} only starts to be sampled with 50 ns, and much longer simulations would be necessary to obtain a complete picture of the dielectric response.
We note however that the dielectric enhancement is present in \ce{CsGeBr3}, and its absence from Figure~\ref{fig:Fig3}(b) is due to the limited simulation length of molecular dynamics.
This example illustrates the importance of dynamics when investigating low-frequency vibrational responses.

From the heating curves, one can also notice that \ce{CsSnBr3} is the only material showing both FE and AFE displacements in the ground state (see SI Fig.~S4). 
As previously discussed, AFE off-centering is nonpolar and is found here to contribute to the low-frequency Raman response, as observed in Figure~\ref{fig:Fig3}(c) where the first phase transition of \ce{CsSnBr3} at 210~K already shows the formation of a Raman central peak. 
This example shows that Raman central peaks do not solely arise from octahedral tilting, but also from other anharmonic modes.

\begin{figure*}
    \centering
    \includegraphics[width=7in]{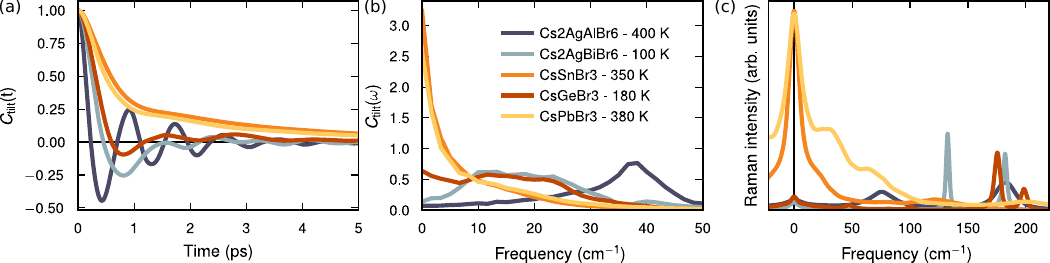}
    \caption{
    Dynamics of the octahedra tilting above the phase transition.
    (a-b) Autocorrelation function of the tilt mode in (a) the time and (b) frequency domains.
    All halides are shown at a temperature 20-30~K above the tilt-ordering transition.
    (c) Normalized Raman spectra at the same temperatures.
    }
    \label{fig:Fig6}
\end{figure*}

Finally, we investigate the dynamics of tilting modes above their related phase transition.
In particular, we look at the tilting autocorrelation function (ACF) $C_\text{tilt}(t)$ for all five halide perovskites at temperatures 20-30~K above their last phase transition involving octahedral tilting.
Results are shown in Figure~\ref{fig:Fig6}(a).
\ce{Cs2AgAlBr6} is used as a reference here, as it exhibits no structural instabilities and retains well-defined phonon modes.
Consequently, its ACF exhibits regular damped oscillations.
From their ACFs, tilting in both \ce{Cs2AgBiBr6} and \ce{CsGeBr3} is underdamped, while it is clearly overdamped in \ce{CsPbBr3} and \ce{CsSnBr3}.
This difference is even more visible in the frequency domain, which is shown in Figure~\ref{fig:Fig6}(b).
In this case, the underdamped modes show clear finite-frequency peaks, while overdamped modes lead to the formation of a central peak.
The tendency toward overdamping is naturally enhanced close to phase transitions, where the frequency of the soft mode $\omega_n$ decreases and becomes smaller than its damping $\gamma$. 
The corresponding spectral weight is then transferred toward zero frequency, producing a central response. 
For strongly damped modes, this overdamped regime ($\omega_n\ll\gamma$) can extend over a broader temperature range on either side of the transition, whereas more weakly damped modes quickly recover a finite-frequency response away from the transition.

These dynamics can be directly compared with the Raman spectra shown in Figure~\ref{fig:Fig6}(c).
For both \ce{CsPbBr3} and \ce{CsSnBr3}, the Raman spectra in the cubic phase are dominated by the central peak, which can be explained by the overdamped tilting modes.
In contrast, the double perovskites and \ce{CsGeBr3} do not show a pronounced central peak, but instead well-defined finite-frequency peaks. 
Note that \ce{CsGeBr3} at 180~K has not yet reached the cubic phase, and the Raman peaks visible in Figure~\ref{fig:Fig6}(c) arise from the reduced symmetry due to B-site off-centering.
The over or underdamping regime therefore explains why some central peaks vanish at high temperatures in Figure~\ref{fig:Fig3}(c), while other materials retain a strong central peak above their phase transition temperature.
This picture connects the assignment of the central response to octahedral tilting with its description in terms of strongly broadened low-frequency Raman excitations \cite{Lim2024_ACSEnerLett}, while also explaining why \ce{CsGeBr3} and \ce{Cs2AgBiBr6} do not retain a strong central peak above the phase transition.
For the latter, experimental measurements already observed well-defined soft modes that approach zero frequency at the transition and disappear in the cubic phase, although no central peak was resolved experimentally \cite{Cohen2022_AdvMater}. 
In our simulations, the central response is confined to a narrow temperature range around the transition, while a finite-frequency response is rapidly recovered below it.
Overdamping has also been identified as the origin of central Raman peaks in \ce{BaZrO3} \cite{Rosander2025_Raman}.
Within this picture, the Raman coupling determines the intensity of a given soft mode, whereas its frequency and damping determine how the associated spectral weight is distributed between finite and near-zero frequencies.
Similarly, slow polar reorientation can shift the associated soft modes to frequencies much lower than commonly probed by phonons.
Together, these results highlight how the dynamics of anharmonic modes influence the low-frequency vibrational spectra, in addition to mode symmetry and IR/Raman couplings.

% ============================================================
\section{Conclusions}

In conclusion, we simulated the infrared and Raman spectra, as well as the dielectric function, over a wide temperature range and for various perovskites, including halide, double halide and oxide perovskites.
By systematically comparing the low-frequency responses with the types of instabilities present in each material, our results show that these responses are closely related to anharmonic modes and their associated phase transitions.
In particular, we find that the response is maximal at the phase transition temperatures.
By comparing infrared and Raman responses, we find a clear correspondence between the symmetry of the underlying instabilities and their dominant spectroscopic response.
Due to their polar nature, ferroelectric B-site off-centering is found to produce soft infrared modes, which in turn give rise to an enhancement of the dielectric function.
In contrast, Raman central peaks can arise from nonpolar anharmonic modes, including tilting of the \ce{BX6} octahedra and anti-ferroelectric off-centering modes.
In addition to this symmetry dependence, we also highlight the importance of the infrared and Raman coupling in the formation of low-frequency features.
For example, \ce{SrTiO3} fails to produce a notable central Raman peak due to its small coupling $\frac{d\chi}{dQ}$.

In addition to the mode symmetry and its infrared or Raman coupling, we also show that dynamics play an important role in the low-frequency responses.
For example, the slow reorientation of B-site off-centerings in \ce{CsSnBr3} and especially in \ce{CsGeBr3} can lead to responses at much lower frequencies than usual phonon modes.
We also find that the damping of octahedra tilting greatly influences the low-frequency Raman response.
While the commonly observed dominant central peak in the cubic phase of \ce{CsPbBr3} and \ce{CsSnBr3} can be associated with overdamped modes, our results show that central peaks are also present in the Raman spectra of \ce{CsGeBr3} and \ce{Cs2AgBiBr6}, although limited to temperatures close to the phase transition.
Overall, these results help explain the presence and the formation of low-frequency features in vibrational spectra, especially for Raman spectroscopy where the interpretation of the central peak remained elusive.
While the current work focuses on perovskite materials, its findings are expected to extend to any material exhibiting strongly anharmonic phonon modes.

% ============================================================
\section{Methods}

\subsection{Vibrational spectra and dielectric functions}

All spectra are computed using Fourier transforms of auto-correlation functions. 
For the Raman spectrum $\tens{I}^\text{R}$, this reads \cite{Thomas2013_Raman}
\begin{equation}
    I^\text{R}_{ij}(\omega) \propto \frac{\omega}{1-e^{-\beta\hbar\omega}}\int\langle\chi_i(\tau)\chi_j(\tau+t)\rangle_\tau\ e^{-i\omega t}dt,
\end{equation}
where $\beta=\frac{1}{k_\text{B}T}$ is the inverse temperature and $\vect{\chi}(t)$ is the dielectric susceptibility at time $t$. 
For simplicity, we only consider the trace of the Raman response, that is $I^\text{R}_{xx}+I^\text{R}_{yy}+I^\text{R}_{zz}$.
Susceptibilities are obtained along the MD trajectory using a tensorial neuroevolution potential (TNEP).
For each material, the Raman spectra are normalized using the maximal intensity of finite-frequency peaks at the lowest temperature.
The central peak is defined as the intensity at zero frequency $\tens{I}^\text{R}(\omega=0)$.
The central peak intensity presented in Figure~\ref{fig:Fig3} therefore corresponds to the ratio between the zero frequency response and the maximal response close to the ground state.

The infrared spectrum $\tens{I}^\text{IR}$ and the dielectric function $\vect{\varepsilon}(\omega)=\vect{\varepsilon}'(\omega)+i\vect{\varepsilon}''(\omega)$ are computed using a charge-informed neuroevolution potential (qNEP) \cite{Fan2026_qnep}. 
The imaginary part of the dielectric function $\vect{\varepsilon}_2(\omega)$ is first calculated as
\begin{equation}
    \varepsilon''_{ij}(\omega) = \frac{\beta}{\varepsilon_0\omega V}\int\langle\dot{P}_i(\tau)\dot{P}_j(\tau+t)\rangle_\tau\ e^{-i\omega t} dt,
\end{equation}
where $\varepsilon_0$ is the vacuum permittivity, $V$ is the volume, and $\dot{\vect{P}}$ is the time derivative of the dipole $\vect{P}$. 
It is obtained from the Born effective charges $\tens{Z}$ as
\begin{equation}
    \dot{\vect{P}} = \frac{d\vect{P}}{dt} = \frac{d\vect{P}}{d\vect{r}}\cdot\frac{d\vect{r}}{dt} = \tens{Z}\cdot \vect{v}.
\end{equation}
The real part of the dielectric function $\vect{\varepsilon}'(\omega)$ is then derived using the Kramers-Kronig relation, and the infrared spectrum is obtained as $\tens{I}_\text{IR}(\omega)=\omega \vect{\varepsilon}''(\omega)$.
Because of the finite-length MD and incomplete sampling of low frequencies, the dielectric constant presented in this work is defined as the real part of the dielectric constant at a frequency of 1~\icm.
Also note that our results only include the vibrational contribution to the dielectric constant, and that the electronic part $\varepsilon_\infty$ is not included.

\subsection{Phonon dispersion curves and mode projection}

Phonon dispersion curves are calculated with the phonopy package \cite{Togo_2015} from force constant matrices obtained using the calorine package \cite{calorine}.
Displacements $\Gamma_4^-$ and R$_4^+$ are extracted from the phonon dispersion curves from \ce{BaTiO3} and \ce{CsPbBr3}, respectively.
The same paths are then used for all materials when plotting Figures~\ref{fig:Fig4}(e-h).

The analysis of the off-centering distortion $\vect{Q}_\text{disp}$ is done following Ref.~\citenum{Fransson2026_Ge}.
For B-site $i$, the displacement $\vect{Q}_\text{disp}^i$ is defined as
\begin{equation}
    \vect{Q}_\text{disp}^i = \vect{r}_i - \frac{1}{6}\sum_{n=1}^6 \vect{r}_n,
\end{equation}
where $\vect{r}_i$ is the position of B-site $i$, and $\vect{r}_n$ are the positions of the 6 neighbouring X-sites.
The tilting angles used for $\vect{Q}_\text{tilt}$ are obtained following the method presented in Ref.~\cite{Baskurt2026_Faraday}.

For a given distortion $\vect{Q}(t)$, its corresponding autocorrelation function $\vect{C}(t)$ is defined as
\begin{equation}
    C_i(t) = \frac{\langle Q_i(\tau)\cdot Q_i(\tau+t)\rangle_\tau}{\langle Q_i^2\rangle}.
\end{equation}
The tilting $\vect{C}_\text{tilt}(t)$ and off-centering $\vect{C}_\text{disp}(t)$ autocorrelation functions are obtained using $\vect{Q}_\text{tilt}(t)$ and $\vect{Q}_\text{disp}(t)$, respectively.
The autocorrelation function in the frequency domain is obtained using a Fourier transform.

\subsection{Neuroevolution potentials qNEP and TNEP}

All machine learning interatomic potentials used in this work are based on the neuroevolution potential (NEP) framework \cite{Fan2021_NEP,Dong2024_NEP,Ying2025_NEP}.
NEP models are trained using the GPUMD package \cite{XuBuPan25_GPUMD}.
Infrared and Raman spectra are simulated using a combination of charge-informed NEP (qNEP) \cite{Fan2026_qnep} and tensorial NEP (TNEP) \cite{Xu2023_TNEP}.
To facilitate training, existing NEP models were used to generate new training data.
The training workflow followed the following steps:
\begin{enumerate}
    \item NEP models are used to perform NPT runs at various temperatures to generate new small structures.
    \item DFPT is performed to obtain $\tens{Z}$ and $\vect{\chi}$ for these new structures.
    \item Energies, forces, and virials of the new structures are obtained using the initial NEP model.
    \item qNEP is trained on energies, forces, virials, and BECs, while TNEP is trained on the dielectric susceptibilities.
\end{enumerate}
qNEP models are trained using both the initial and additional structures.
The energy, forces, and virial are predicted using the initial NEP model to preserve the potential energy surface while adding BECs via the qNEP formalism.
Additionally, this distillation scheme considerably reduces the computational workload.
We find that the qNEP models trained this way correctly retained the phase diagrams and transition temperatures of the initial NEP model (see details in the SI table~S3).
Note that for \ce{BaTiO3}, we used the existing qNEP from Ref.~\cite{Fan2026_qnep}, and only trained a new TNEP model.
All models were trained for $300\times10^3$ SNES generations using the GPUMD package.
Other hyperparameters were set to match those of the initial NEP models, and therefore vary considerably between materials.
DFPT calculations were all performed using the Vienna \textit{ab initio} simulation package (VASP) \cite{Kresse_1996_1,Kresse_1996_2}.
Because DFPT parameters vary between materials, detailed information regarding the choice of functional, the reciprocal space grid, and the cutoff energy is given in the SI Table~S2.
%All qNEP and TNEP models trained and used in this work are freely available at Ref.~\cite{}.

\subsection{Molecular dynamics simulations}

The resulting qNEP models were first used to perform the heating runs.
All molecular dynamics calculations were performed using the GPUMD package \cite{XuBuPan25_GPUMD}.
For conventional \ce{ABX3} perovskites, $20\times20\times20$ supercells were used, corresponding to structures containing 40,000 atoms.
The same number of atoms was also used for double perovskites.
The time step was fixed to 1 fs.
The heating rate was set to 10~K/ns for all perovskites, and a MTTK barostat was used to simulate the NPT ensemble \cite{Martyna1994_MTTK}.
Results of the heating runs for all materials are presented in the SI.
The lattice constants obtained from heating were then used to create new structures, which were used to obtain spectra.
Production runs for spectra were performed in the NVT ensemble using a Nosé-Hoover chain thermostat \cite{Martyna1992_NHC}.
Equilibration of 1 ns was performed before the 2 ns production, where the BECs and dielectric susceptibilities were sampled every 20 fs.
For the low-frequency spectra presented in Figure~\ref{fig:Fig5}, we performed 50 ns trajectories with the same 20 fs sampling rate.
The off-centering and tilting ACFs presented in Figure~\ref{fig:Fig5} and \ref{fig:Fig6} were obtained through shorter MD of 200 ps and 20 ps in the NVT ensemble, with positions sampled every 50 fs and 5 fs, respectively.

% ============================================================

\section*{Acknowledgments}
This work was supported by the Swedish Strategic Research Foundation through a Future Research Leader programme (FFL21-0129), the European Research Council (ERC Starting Grant no. 101162195), and the Knut and Alice Wallenberg Foundation (Nos.~2023.0032). The computations were enabled by resources provided by the National Academic Infrastructure for Supercomputing in Sweden (NAISS), partially funded by the Swedish Research Council through grant agreement no. 2022-06725.

The authors used AI tools (ChatGPT, OpenAI) during the preparation of this manuscript for assistance with writing and language improvements.
All scientific content, interpretation, and final wording were reviewed and approved by the authors.

\bibliography{BIB}

\end{document}

\title{Supporting Information: \\Unifying Low-Frequency Raman and Infrared Signatures across Perovskites}

\author{Ethan Berger}
\email{ethan.berger@chalmers.se}
\affiliation{Department of Physics and Astronomy, Chalmers University of Technology, SE-41296 Gothenburg, Sweden}
\author{Erik Fransson}
\affiliation{Department of Physics and Astronomy, Chalmers University of Technology, SE-41296 Gothenburg, Sweden}
\author{Apinya Ngoipala}
\affiliation{Department of Physics and Astronomy, Chalmers University of Technology, SE-41296 Gothenburg, Sweden}
\author{Lucas Svensson}
\affiliation{Department of Physics and Astronomy, Chalmers University of Technology, SE-41296 Gothenburg, Sweden}
\author{Priyanka Deswal}
\affiliation{Department of Physics and Astronomy, Chalmers University of Technology, SE-41296 Gothenburg, Sweden}
\author{Mehmet Baskurt}
\affiliation{Department of Physics and Astronomy, Chalmers University of Technology, SE-41296 Gothenburg, Sweden}
\author{Paul Erhart}
\affiliation{Department of Physics and Astronomy, Chalmers University of Technology, SE-41296 Gothenburg, Sweden}
\author{Julia Wiktor}
\email{julia.wiktor@chalmers.se}
\affiliation{Department of Physics and Astronomy, Chalmers University of Technology, SE-41296 Gothenburg, Sweden}

\maketitle

\section{Training of qNEP and TNEP models}

All machine learning potentials used in this work are based on neuroevolution potentials (NEP) \cite{Ying2025_NEP}.
In particular, we either use the tensorial NEP (TNEP) \cite{Xu2023_TNEP}, or a charge informed NEP (qNEP) \cite{Fan2026_qnep}
All models are trained using the GPUMD package \cite{XuBuPan25_GPUMD}, which is consequently also used for molecular dynamics.
To reduce the computational cost of training, qNEP models were distilled from existing reliable NEP models.
The initial training sets were augmented by adding structures sampled at various temperatures.
The energy, forces, and virial of these additional structures were obtained using the initial NEP models, which could be referred to as self-distillation.
The BECs used for qNEP and the dielectric susceptibility used for TNEP were obtained through density functional perturbation theory (DFPT) calculations.
While qNEP was trained using both the initial and additional structures, only the latter were used for TNEP models.
Information about the functional and number of structures used for the initial NEP models, as well as the sampled temperatures and the number of additional structures are given in table~\ref{tab:Training}.
Note that for \ce{BaTiO3}, we reused the qNEP model from Ref.~\cite{Fan2026_qnep}, and the TNEP model was consequently trained using existing DFPT calculations.

\begin{table}[h]
    \begin{tabular}{c|c|c|c|c}
         Material        & Functional & Initial structures & Sampled temperatures (K) & Additional structures \\
         \hline
         \ce{Cs2AgAlBr6} & SCAN+rVV10 & 550 & 50, 100, 150, 200, 250, 300, 350, 400, 450 &  900\\
         \ce{Cs2AgBiBr6} & SCAN+rVV10 & 614 &  50, 100, 150, 200, 250, 300, 350, 400, 450, 500, 550, 600 & 1200\\
         \ce{CsPbBr3}    & vdW-DF-cx  & 618 &   50, 100, 150, 200, 250, 300, 350, 400, 450, 500, 550, 600, 650, 700 & 1400\\
         \ce{CsSnBr3}    & r2SCAN50   & 103 & 50, 100, 150, 200, 250, 300, 350, 400 &  754\\
         \ce{CsGeBr3}    & r2SCAN50   & 110 & 10--700&  520\\
         \ce{SrTiO3}     & r2SCAN     & 832 & 50, 100, 150, 200, 250, 300, 350, 400 & 1200\\
    \end{tabular}
    \caption{Creation of new training data. The initial functional, initial number of structures/atoms, the sampled temperatures and the number of additional structures/atoms are given for all materials.}
    \label{tab:Training}
\end{table}

Parity plot for the energy, forces, virial, BECs, and dielectric susceptibility, along with their respective RMSEs, can be found in Figures~\ref{fig:SI_AgAl}--\ref{fig:SI_STO}, each figure showing a different material.

\clearpage

\section{Density functional perturbation theory calculations}

All DFT calculations were performed using the Vienna ab initio packages (VASP). 
Density functional perturbation theory (DFPT) was used to obtain the dielectric susceptibility and the Born effective charges, which were in turn used to train TNEP and qNEP models.
Given that the DFT parameters varied between materials, the following table~\ref{tab:DFPT} summarizes the important parameters for each material.
Note that in most cases, the initial NEP models were trained on meta-GGA functionals, which currently cannot be used in DFPT calculations.
This motivates the use of a different functional for DFPT calculations.

\begin{table}[h]
    \begin{tabular}{c|c|c|c|c|c}
         Material        & Atoms & Kpoints           & Functional & $E_\text{cut}$ (eV) & $E_\text{diff}$ (eV) \\
         \hline
         \ce{Cs2AgAlBr6} & 20    & $8\times8\times8$ & PBEsol     & 520        & $10^{-6}$  \\
         \ce{Cs2AgBiBr6} & 20    & $8\times8\times8$ & PBEsol     & 520        & $10^{-6}$  \\
         \ce{CsPbBr3}    & 20    & $8\times8\times8$ & vdW-DF-cx  & 300        & $10^{-6}$  \\
         \ce{CsSnBr3}    & 40    & $4\times4\times4$ & PBEsol     & 520        & $10^{-8}$  \\
         \ce{CsGeBr3}    & 20    & $8\times8\times8$ & PBEsol     & 500        & $10^{-8}$  \\
         \ce{BaTiO3}     & 20    & $8\times8\times8$ & PBEsol     & 500        & $10^{-8}$  \\
         \ce{SrTiO3}     & 20    & $8\times8\times8$ & PBEsol     & 520        & $10^{-8}$  \\
    \end{tabular}
    \caption{Parameters used for DFPT calculations. 
    The number of atoms, density of the reciprocal space mesh, choice of functional, cutoff energy ($E_\text{cut}$), and the convergence energy ($E_\text{diff}$) are given for all 7 perovskites.}
    \label{tab:DFPT}
\end{table}

\clearpage

\section{Heating runs and phase transition temperatures}

\begin{figure}
    \centering
    \includegraphics[width=7in]{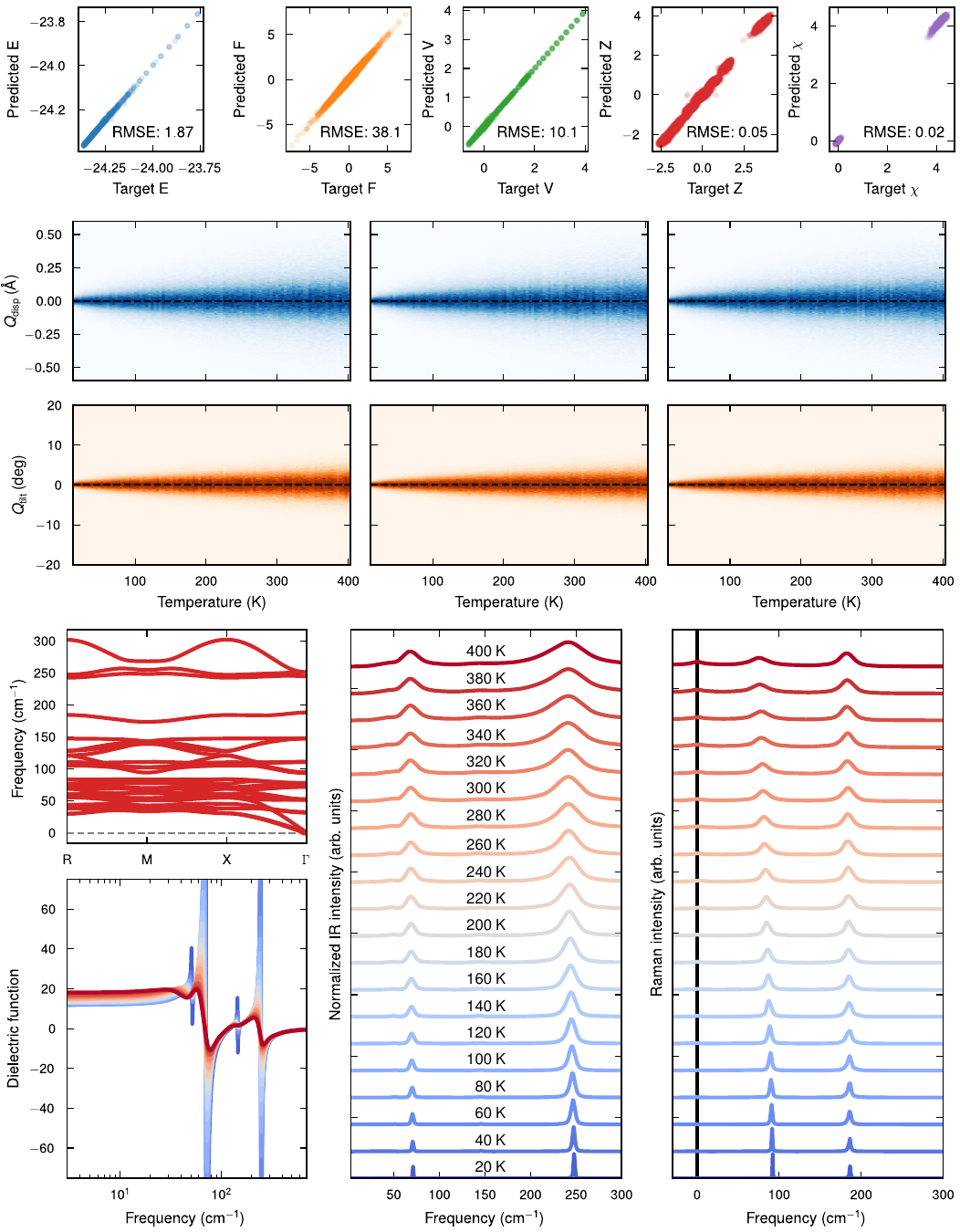}
    \caption{Parity plots, heating curves, phonon dispersion curves, dielectric function, IR spectra, and Raman spectra of \ce{Cs2AgAlBr6}.}
    \label{fig:SI_AgAl}
\end{figure}

\begin{figure}
    \centering
    \includegraphics[width=7in]{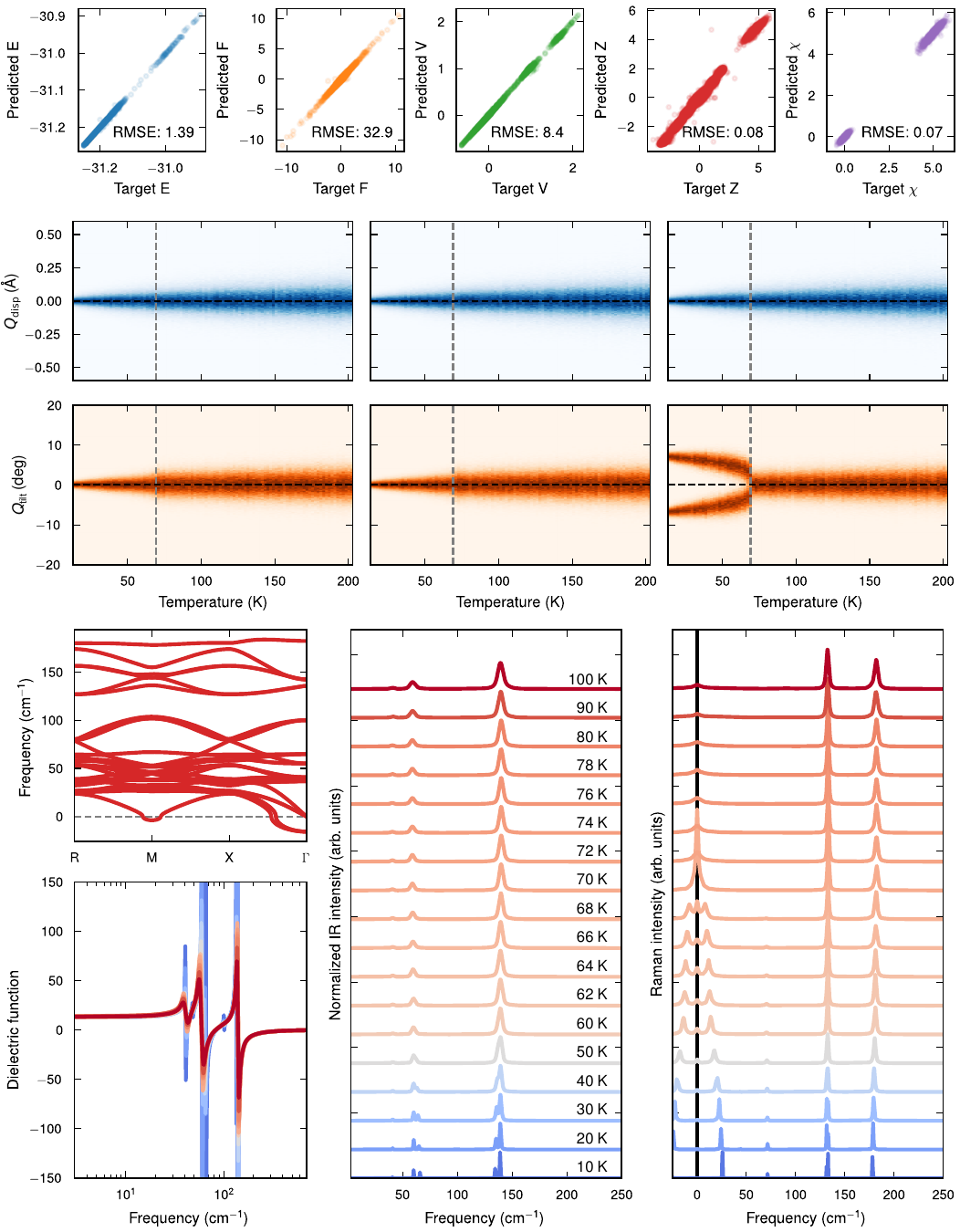}
    \caption{Parity plots, heating curves, phonon dispersion curves, dielectric function, IR spectra, and Raman spectra of \ce{Cs2AgBiBr6}.}
    \label{fig:SI_AgBi}
\end{figure}

\begin{figure}
    \centering
    \includegraphics[width=7in]{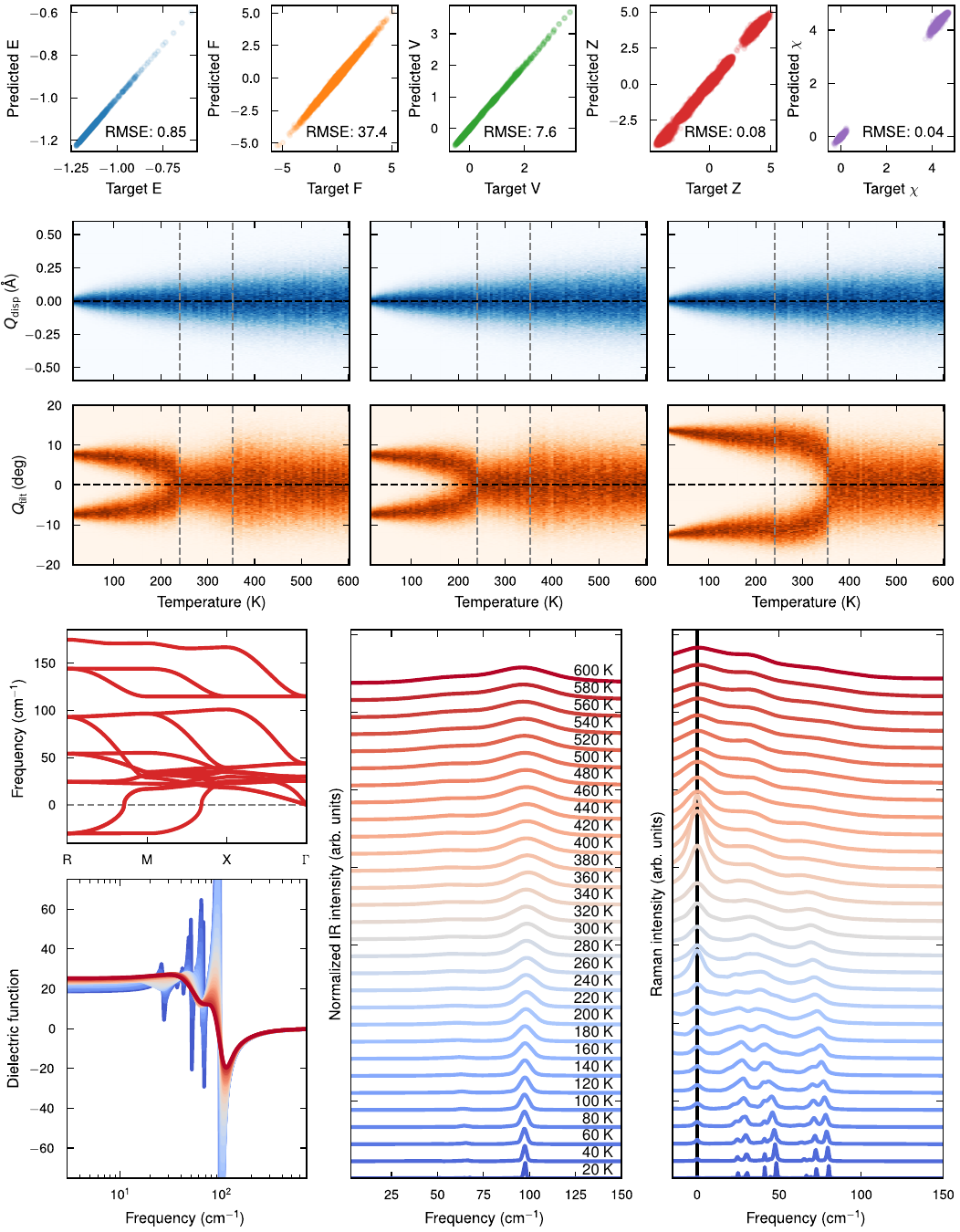}
    \caption{Parity plots, heating curves, phonon dispersion curves, dielectric function, IR spectra, and Raman spectra of \ce{CsPbBr3}.}
    \label{fig:SI_Pb}
\end{figure}

\begin{figure}
    \centering
    \includegraphics[width=7in]{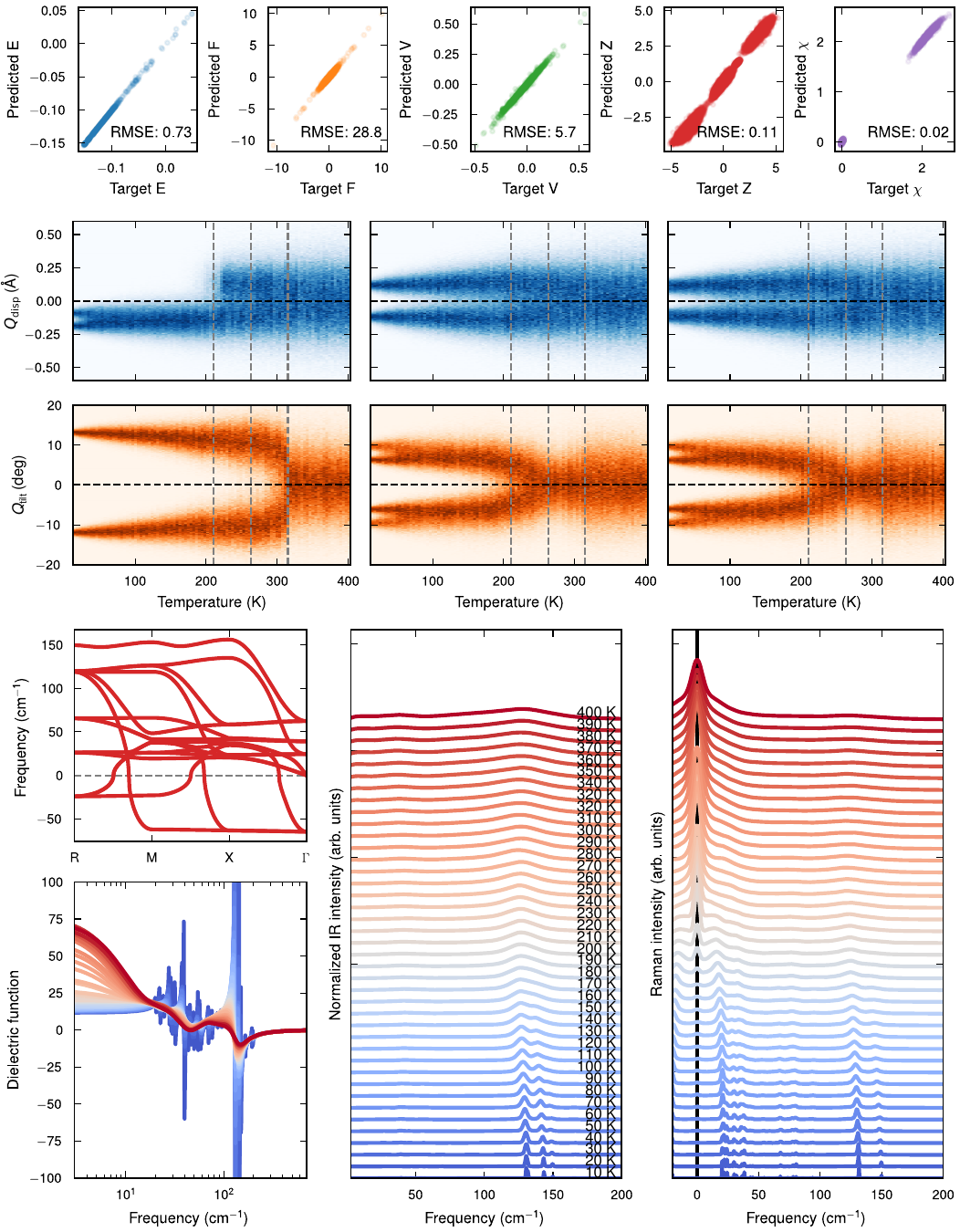}
    \caption{Parity plots, heating curves, phonon dispersion curves, dielectric function, IR spectra, and Raman spectra of \ce{CsSnBr3}.}
    \label{fig:SI_Sn}
\end{figure}

\begin{figure}
    \centering
    \includegraphics[width=7in]{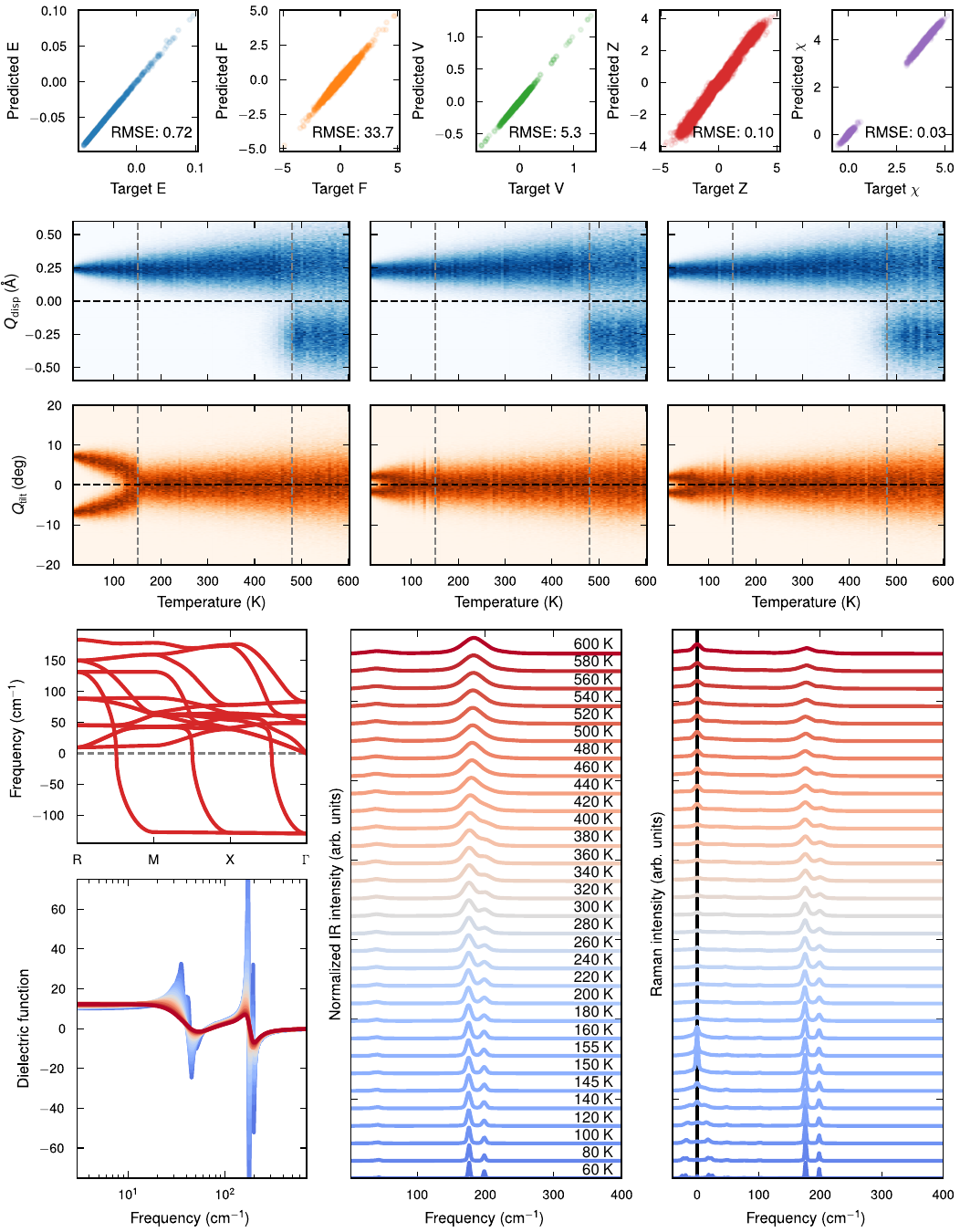}
    \caption{Parity plots, heating curves, phonon dispersion curves, dielectric function, IR spectra, and Raman spectra of \ce{CsGeBr3}.}
    \label{fig:SI_Ge}
\end{figure}

\begin{figure}
    \centering
    \includegraphics[width=7in]{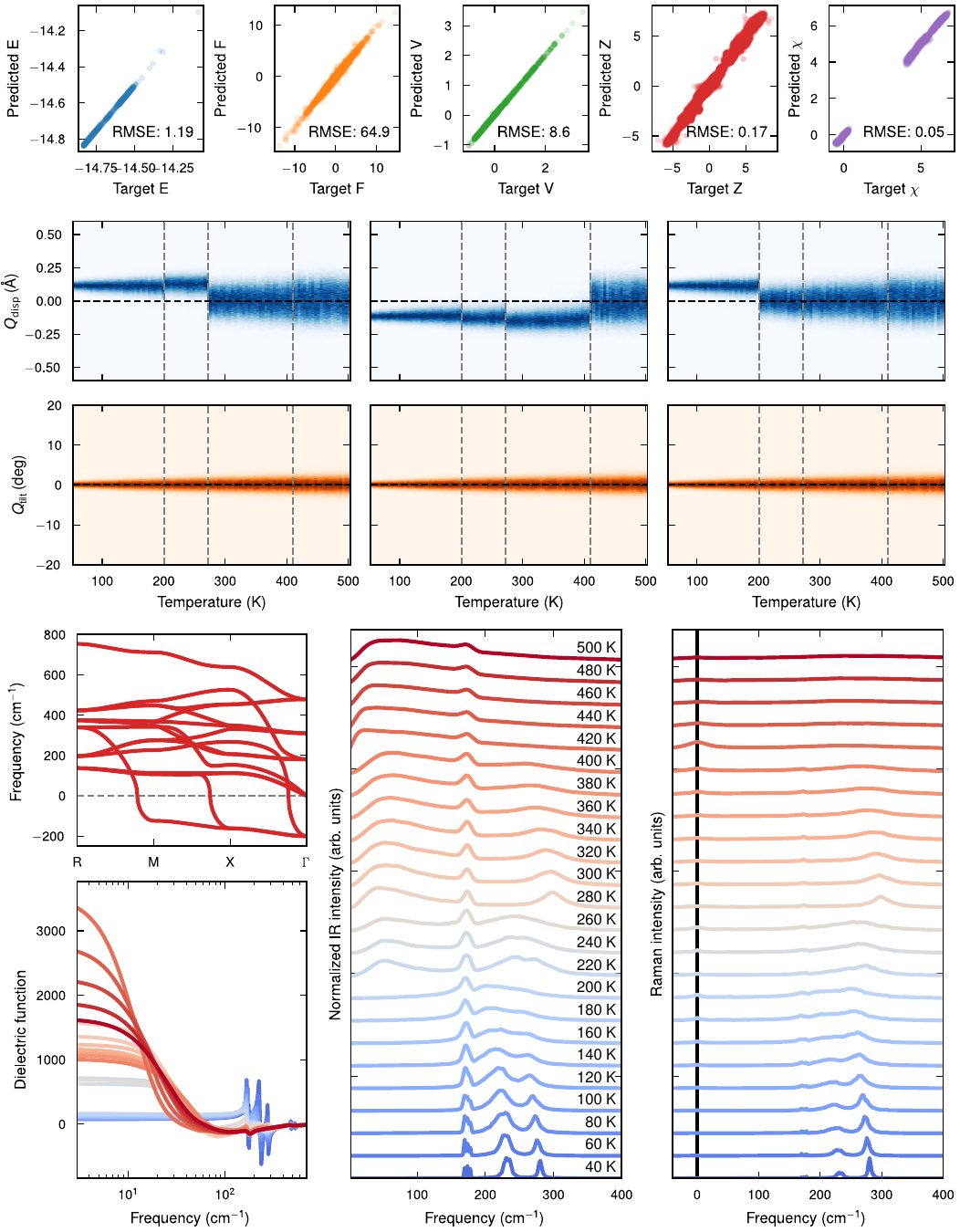}
    \caption{Parity plots, heating curves, phonon dispersion curves, dielectric function, IR spectra, and Raman spectra of \ce{BaTiO3}.}
    \label{fig:SI_BTO}
\end{figure}

\begin{figure}
    \centering
    \includegraphics[width=7in]{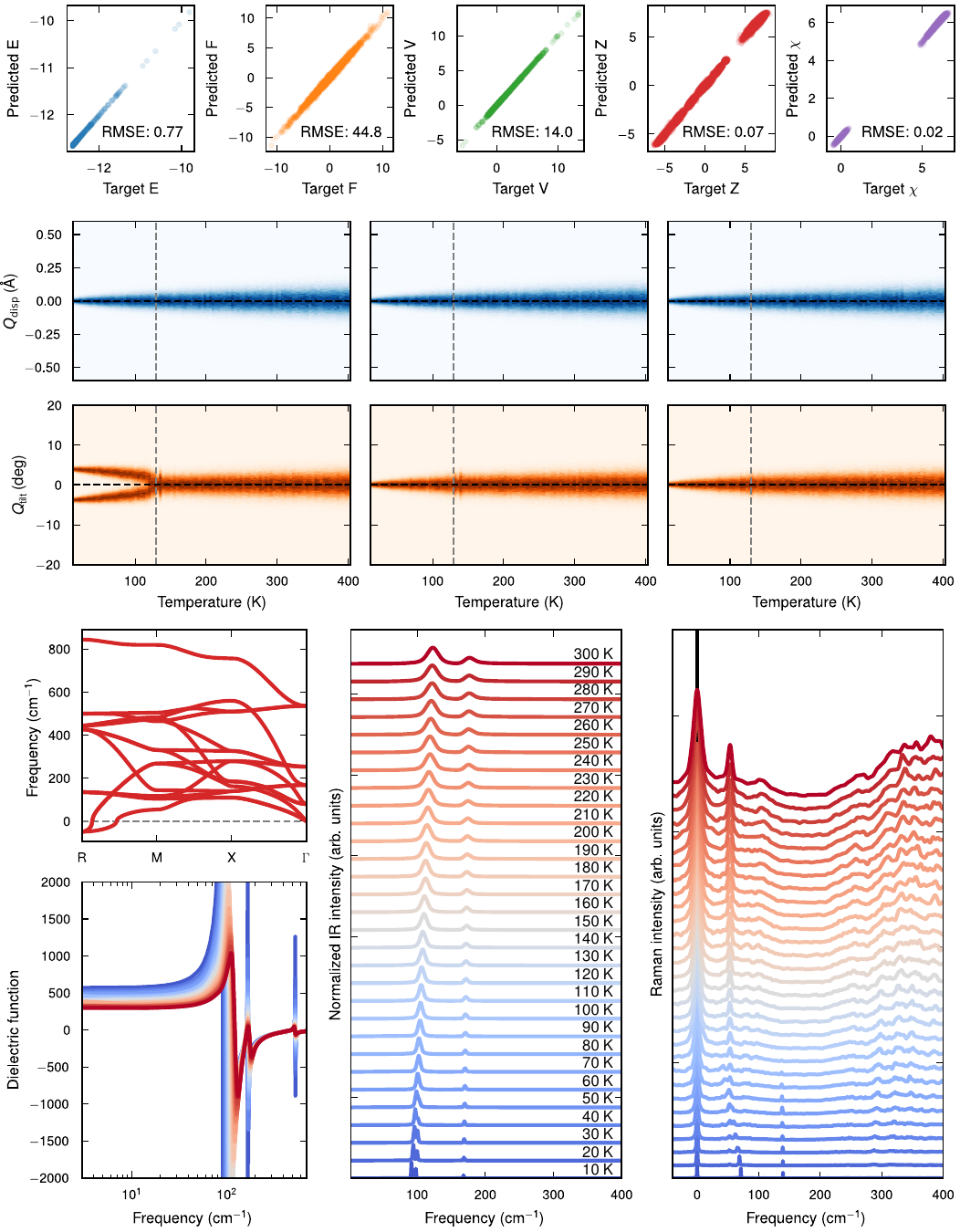}
    \caption{Parity plots, heating curves, phonon dispersion curves, dielectric function, IR spectra, and Raman spectra of \ce{SrTiO3}.}
    \label{fig:SI_STO}
\end{figure}

Phase transitions were obtained through heating runs.
The resulting $Q_\text{disp}$ and $Q_\text{tilt}$ along the heating trajectories are reported in all three directions in Figures~\ref{fig:SI_AgAl}--\ref{fig:SI_STO}.
Phase transitions are reported in table~\ref{tab:Tc}, and compared with values from previous NEP models as well as experimental measurements.
Overall, the distilled qNEPs correctly reproduce the phase transitions from the initial NEP models, with variations of 20--40 K.
In some cases, the difference can be explained by different supercell sizes or heating rates, or by hysteresis.
More importantly, the qNEP models correctly predict phase diagrams, which are essential for the current work.

\begin{table}[h]
    \begin{tabular}{c|c|c|c}
         Material        & This work qNEP & Initial NEP & Experiment \\
         \hline
         \ce{Cs2AgAlBr6}      & None          & None \cite{Baskurt2026_Faraday} & -- \\
         \ce{Cs2AgBiBr6}      & 69            & 50 \cite{Baskurt2026_Faraday} & 120 \cite{Tower_Tc_AgBi} \\
         \ce{CsPbBr3}         & 240, 354      & 276, 365 \cite{Wiktor2023_CsPbBr3} & 361, 403 \cite{Stoumpos2013_CsPbBr3} \\
         \ce{CsSnBr3}         & 210, 263, 315 & \cite{Fransson2026_Ge} & 247, 274, 292 \cite{Mori1986_Tc_CsSnBr3} \\
         \ce{CsGeBr3}         & 150,480       & 110, 429 \cite{Deswal2026_Ge} & --, 511 \cite{Thiele1987_CsGeBr3}  \\
         \ce{BaTiO3}          & 201, 272, 410 & 151, 235, 390 \cite{Fan2026_qnep} & 183, 278, 393 \cite{Merz1949_BTO} \\
         \ce{SrTiO3}          & 125           &       --      & 105 \cite{Guennou2010_STO} \\
         \ce{SrTiO3} (-7 GPa) & 53, 114, 248  &       --      & -- \\
    \end{tabular}
    \caption{Phase transition temperatures for every materials.
    Results from this work are compared with previous NEP models and experimental measurements.}
    \label{tab:Tc}
\end{table}

\begin{figure}
    \centering
    \includegraphics[width=7in]{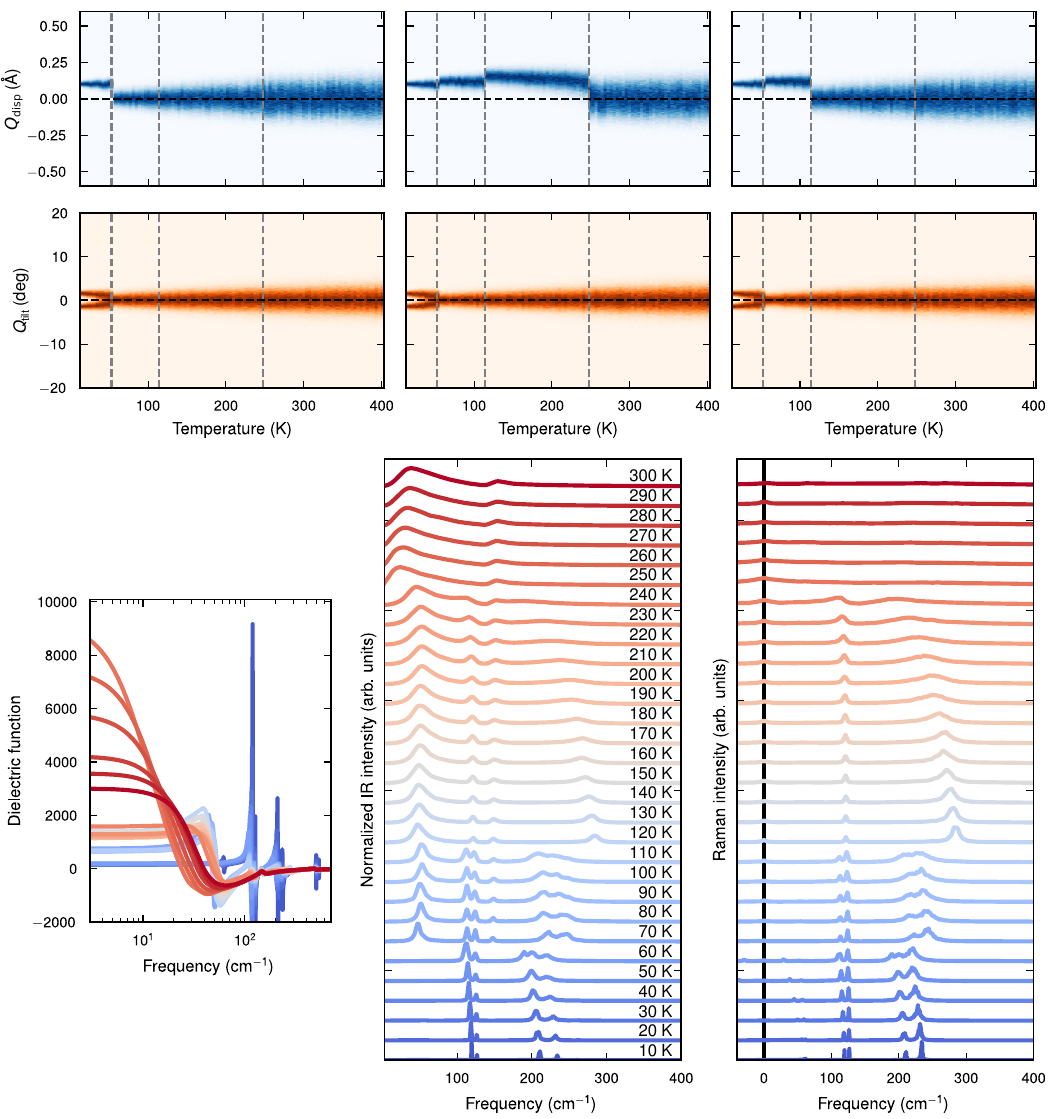}
    \caption{Heating curves, dielectric function, IR spectra, and Raman spectra of \ce{SrTiO3} under negative pressure of -7 GPa.}
    \label{fig:SI_STOP}
\end{figure}

\begin{figure}
    \centering
    \includegraphics[width=7in]{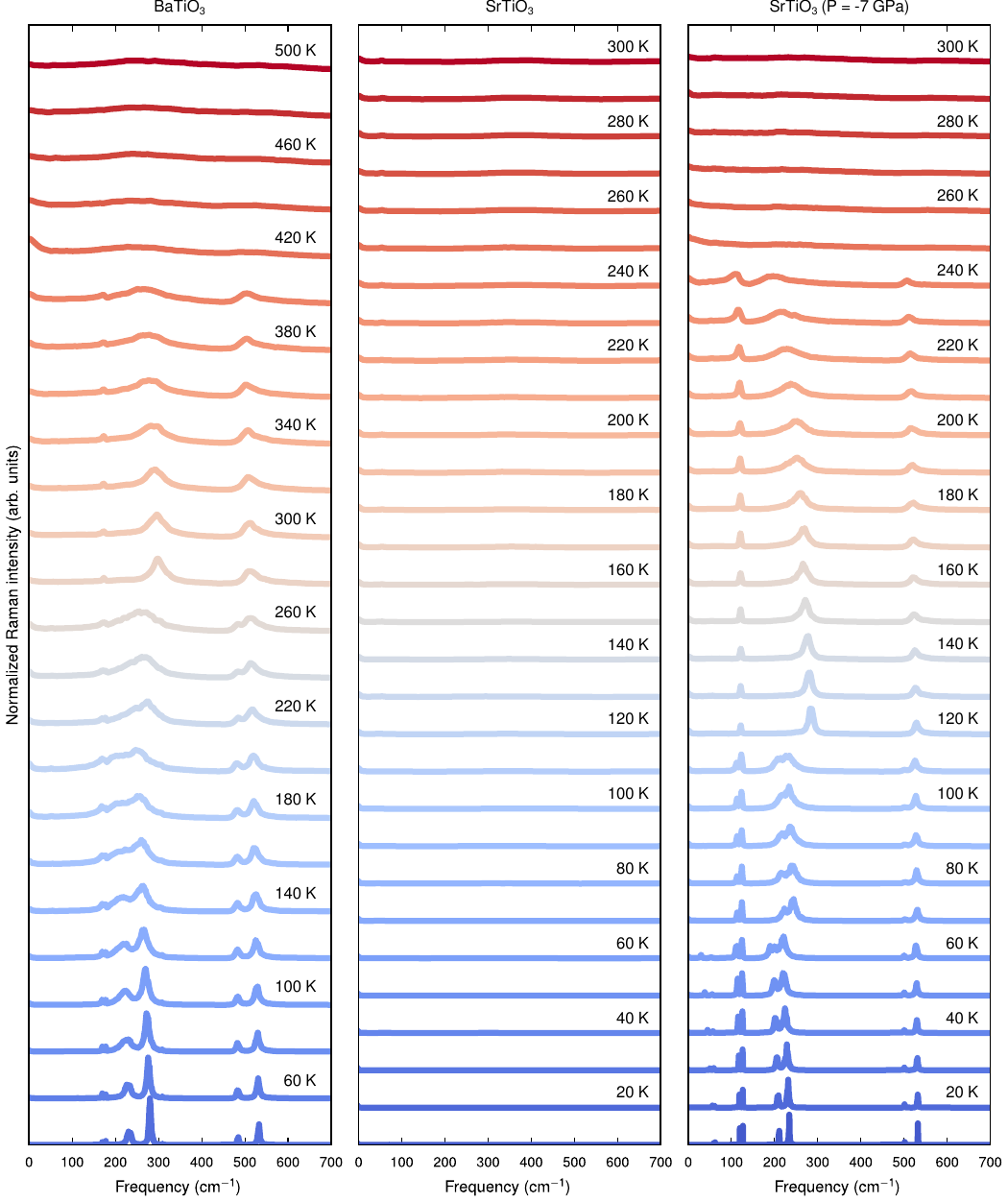}
    \caption{Raman spectra of \ce{BaTiO3} and \ce{SrTiO3} at both pressures.}
    \label{fig:SI_norm}
\end{figure}

\begin{figure}
    \centering
    \includegraphics[width=5.6in]{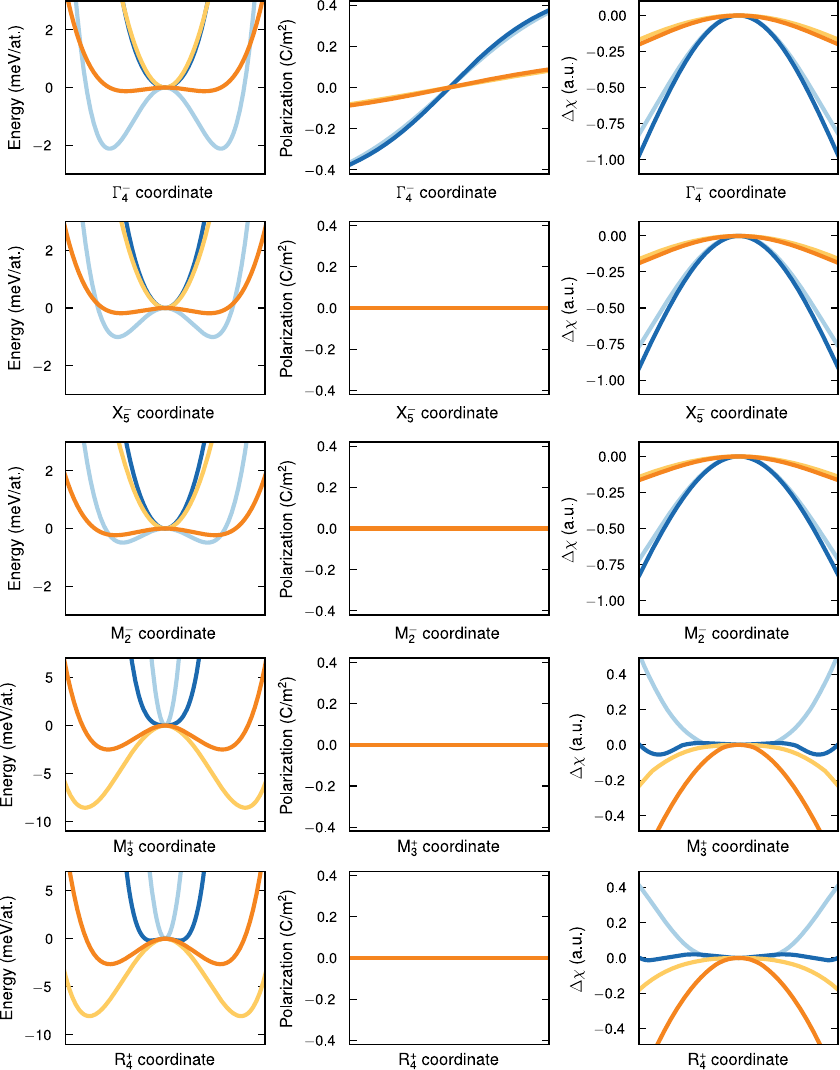}
    \caption{Energy, polarization and dielectric susceptibility variation along all 5 anharmonic phonon modes.
    $\Gamma_4^-$, $X_5^-$, and $M_2^-$ correspond to the three imaginary off-centering modes of \ce{BaTiO3}.
    $M_3^+$ and $\text{R}_4^+$ correspond to the two imaginary tilting modes of \ce{CsPbBr3}.}
    \label{fig:SI_modes}
\end{figure}
\clearpage
\bibliography{BIB}